\documentclass[twocolumn,superscriptaddress,nofootinbib,floatfix,aps,prb,citeautoscript,longbibliography,10pt]{revtex4-2}
\usepackage[utf8]{inputenc}
\usepackage[T1]{fontenc}
\usepackage{babel}

\usepackage{lineno}
\usepackage{graphicx}
\usepackage{color}
\usepackage{array}
\usepackage{dcolumn}
\usepackage{bm}
\usepackage{multirow}
\usepackage[version=4]{mhchem}
\usepackage{booktabs}
\usepackage{hyperref}
\usepackage{cleveref}

\newcommand{\alex}{\textsc{Alexandria}}

\newcommand{\ehull}{E$_{\mathrm{hull}}$}

\begin{document}

\newcommand{\bochum}{Research Center Future Energy Materials and Systems of the University Alliance Ruhr and ICAMS, Ruhr University Bochum, Universitätsstraße 150, D-44801 Bochum, Germany}
\newcommand{\virginia}{Department of Physics, West Virginia University, Morgantown, WV 26506, USA}
\newcommand{\coimbra}{CFisUC, Department of Physics, University of Coimbra, Rua Larga, 3004-516 Coimbra, Portugal}
\newcommand{\ETH}{Department of Materials, ETH Zürich, Zürich, CH-8093, Switzerland}
\newcommand{\icams}{ICAMS, Ruhr-Universität Bochum, Universitätstrasse 150, 44801 Bochum, Germany and ACEworks GmbH, Hagen-Hof-Weg 1, 44797 Bochum, Germany}

\author{Th\'{e}o Cavignac}
\email{theo.cavignac@rub.de}
\affiliation{\bochum}
\author{Jonathan Schmidt}
\affiliation{\ETH}
\author{Pierre-Paul De Breuck}
\affiliation{\bochum}
\author{Antoine Loew}
\affiliation{\bochum}
\author{Tiago F. T. Cerqueira}
\affiliation{\coimbra}
\author{Hai-Chen Wang}
\affiliation{\bochum}
\author{Anton Bochkarev}
\affiliation{\icams}
\author{Yury Lysogorskiy}
\affiliation{\icams}
\author{Aldo H. Romero}
\affiliation{\virginia}
\author{Ralf Drautz}
\affiliation{\icams}
\author{Silvana Botti}
\email{silvana.botti@rub.de}
\affiliation{\bochum}
\author{Miguel A. L. Marques}
\email{miguel.marques@rub.de}
\affiliation{\bochum}

\date{\today}

\title{AI-Driven Expansion and Application of the Alexandria Database}

\begin{abstract}
We present a novel multi-stage workflow for computational materials discovery that achieves a 99\% success rate in identifying compounds within 100~meV/atom of thermodynamic stability, with a threefold improvement over previous approaches. By combining the \textsc{Matra-Genoa} generative model, \textsc{Orb-v2} universal machine learning interatomic potential, and \textsc{ALIGNN} graph neural network for energy prediction, we generated 119 million candidate structures and added 1.3 million DFT-validated compounds to the \alex{} database, including 74 thousand new stable materials.
The expanded \alex{} database now contains 5.8 million structures with 175 thousand compounds on the convex hull. Predicted structural disorder rates (37--43\%) match experimental databases, unlike other recent AI-generated datasets. Analysis reveals fundamental patterns in space group distributions, coordination environments, and phase stability networks, including sub-linear scaling of convex hull connectivity.
We release the complete dataset, including sAlex25 with 14 million out-of-equilibrium structures containing forces and stresses for training universal force fields. We demonstrate that fine-tuning a GRACE model on this data improves benchmark accuracy. All data, models, and workflows are freely available under Creative Commons licenses.
\end{abstract}

\maketitle

\section{Introduction}

Recent advances in availability of large materials databases serves as evidence of the success of high-throughput (HT) materials discovery~\cite{gnome, wolvertonht,alexandria,hcgat,roadmap, ocpbook, https://doi.org/10.48550/arxiv.2411.11783, Sriram2024} and provide reservoirs of hypothetical compounds that are thermodynamically stable or close to stability. Despite these advances, sampling the vast combinatorial space of possible materials remains computationally intensive and unfeasible by brute force. Fortunately, the development of machine learning (ML) models has greatly accelerated this process~\cite{cgat, orb, mace, m3gnet, bochkarevGraphAtomicCluster2024}. Conversely, we have also witnessed a dramatic increase of data available to train new models~\cite{alexandria, omat24, https://doi.org/10.48550/arxiv.2411.11783, Sriram2024}. Such a positive feedback loop has significantly accelerated the material discovery for technological applications ranging from energy storage to catalysis and electronics.

The \alex{} database represents one of the largest collections of \textit{ab initio} calculations and is the largest open database for thermodynamically stable materials. Currently, it encompasses approximately 5.8 million structures calculated using density functional theory (DFT), of which 175 thousand structures lie on the convex hull.

The previous iteration of this dataset has been extensively utilized by the scientific community.
For instance, all universal force-field models within the top five~\cite{fu2025learningsmoothexpressiveinteratomic,omat24,kim2024dataefficientmultifidelitytraininghighfidelity,orb,zeng2025deepmdkitv3multiplebackendframework,lysogorskiyGraphAtomicCluster2025} of the Matbench Discovery ranking~\cite{Matbench} were trained using \alex{} data.
Some of these force fields have advanced to the point where they can reliably predict structural, vibrational, and thermal properties~\cite{loew_universal_2025}, defect energies~\cite{eastman_openmm_2023,sharma_accelerating_2025}, and infrared spectra~\cite{bhatia_mace4ir_2025}, thereby unlocking new avenues for exploring physical phenomena.

Beyond training machine-learning force fields, \alex{} has also been employed in the development of generative models~\cite{breuck_generative_2025}. In addition to serving as a training resource, this extensive database provides a valuable foundation for data-driven design of functional materials. For instance, subsets of \alex{} have been used to identify novel dielectric semiconductors~\cite{PhysRevMaterials.8.L122201}, novel two dimensional materials~\cite{2ddatabase} and hard magnets~\cite{vishina_stable_2023}. Furthermore, state-of-the-art machine-learning–assisted high-throughput design frameworks for conventional high-$T_c$ superconductors leverage \alex{} as their primary search space~\cite{Cerqueira2023,Cerqueira2024,daSilva2025}.

A major challenge in the growth of \alex{} is to generate novel structures that are potentially located near the convex hull with high success rate.
During standard prototype-based high-throughput studies the success rate is of the order of 0.1\%~\cite{schmidt2017}.
To increase the success rate of standard HT search, Wang \textit{et al.}~\cite{wang2021predicting} used data-driven chemical similarity measures to guide systematic substitution, i.e. replacing elements in stable or near-stable compounds with similar elements.
This strategy significantly improved the success rate up to 9\%, however, the majority of the calculated structures were still far beyond the (meta-)stable threshold.

Since then, the size of openly available convex hull databases has more than doubled~\cite{hcgat,alexandria}, opening up opportunities for training modern ML models to further accelerate materials discovery workflows. Recently, several large-scale computational discovery efforts have substantially expanded the landscape of predicted stable materials. In this work, we present the optimized workflow currently employed in the \alex{} database to systematically generate new stable and meta-stable compounds with enhanced success rates. We adopt a historical perspective, describing our attempts to improve the discovery pipeline, many of which have been enabled by the continuous development of increasingly sophisticated ML models. Our approach builds upon previous strategies while incorporating novel algorithmic improvements and leveraging the latest advances in graph neural networks and generative models for crystal structure prediction~\cite{breuck_generative_2025}.

Google Deepmind's GNoME workflow~\cite{merchant_scaling_2023} combines symmetry-aware substitutions and \textit{ab initio} random structure searching with graph neural networks in an active-learning framework, generating a vast catalog of predicted stable crystals. This approach demonstrates the power of discriminative models operating at scale, particularly for exploring compositionally complex materials. Our work takes a complementary approach, leveraging recent advances in generative models and universal machine learning interatomic potentials to achieve higher success rates while maintaining complete data openness.

A key distinction in methodology relates to compositional complexity and structural order. The GNoME workflow predominantly targets quaternary and quinary compositions, where over 80\% of resulting compounds are predicted to be substitutionally disordered~\cite{disorder}. In contrast, our approach emphasizes discovering ordered crystal structures across all composition ranges, as evidenced by predicted disorder rates (37--43\%) comparable to experimental databases like ICSD. This difference reflects distinct but complementary strategies: exploring compositionally rich phase spaces versus systematically mapping ordered compounds that are more readily synthesizable and characterizable.

Furthermore, we address a critical need in the materials discovery community by making our entire workflow completely open source and releasing all generated data (not just compounds on the convex hull) under permissive licenses that enable unrestricted use for research, including model training and further development. This commitment ensures that the broader community can not only benefit from our discoveries but also is able to reproduce our results by following identical strategies, contribute to continuous improvement of discovery workflows, and build upon our data without legal restrictions. Such transparency is essential for creating a truly collaborative ecosystem for accelerated materials design.

\section{Results and Discussion}

\subsection{Workflow}

\begin{figure}[htb]
    \centering
    \includegraphics[width=.98\columnwidth]{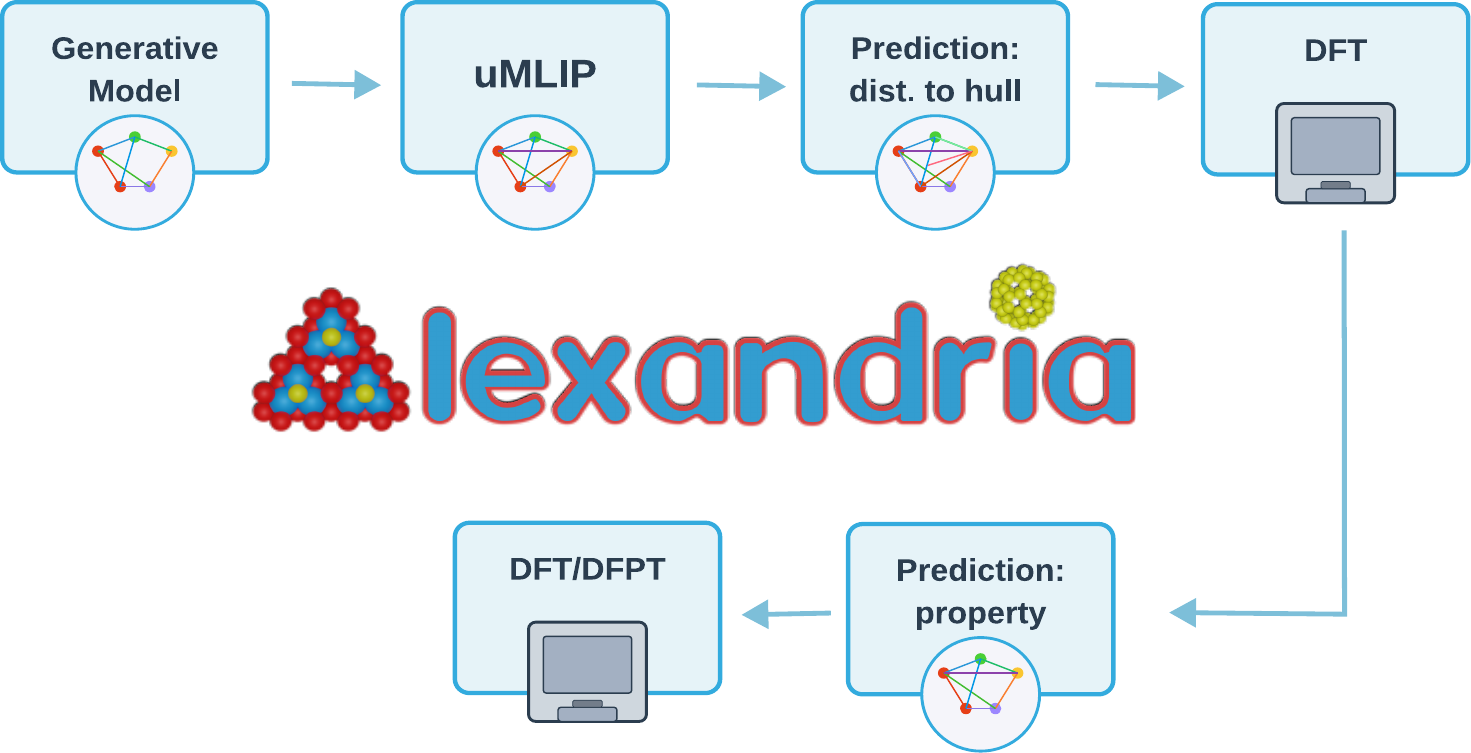}
    \caption{Materials discovery workflow used in \alex{}}
    \label{fig:alexworkflow}
\end{figure}

In \Cref{fig:alexworkflow} we depict the typical workflow used in \alex{}. The cycle works as follows: a model (or a human) proposes compounds composed of a chemical composition and a crystal structure. At this stage, the structure is typically not in dynamical equilibrium.
The next step is a preliminary geometry relaxation with a universal machine learning interatomic potential (uMLIP), that brings the compound close to equilibrium.
In spite of the high quality of modern uMLIPs, we find it advantageous to estimate the distance to the convex hull using a specialized model.
There are two reasons for this, first it is considerably faster to perform model inference than to calculate distances to the hull based on total energies and second it is also more accurate.
At this point we filter the candidate compounds based on their estimated distance to the hull by using a cutoff.
Typical values of this cutoff are between 25~meV/atom and 200~meV/atom, depending on the material class we are investigating and the availability of computer time for the subsequent step.
Selected compounds are then relaxed using DFT and inserted in \alex{} (see \Cref{sec:dft}).

The last steps of the workflow regard specific material properties that we may be interested in.
The workflow proceeds in a similar way, specifically we use purpose-built ML models to provide an estimate of the property at hand (based on either the DFT relaxed structure or the uMLIP one), and if this approximation passes a threshold the property is accurately computed through \textit{ab initio} methods (DFT, density-functional perturbation theory, many-body methods, etc.).
An example of application of this process is our high-throughput study of superconductors~\cite{ssuperconductors}.

During the past few years, we have used a number of different possibilities for each one of the steps that we describe now in more detail.
For the initial generation of the structures many different approaches can be used:
\begin{itemize}
    \item \textbf{Exhaustive enumeration}. This method simply enumerates all possible chemical compositions compatible with a given prototype structure. In spite of the very unfavorable scaling with number of chemical elements, this is a viable approach at least up to ternary compounds. This is the approach used, for example, in Ref.~\cite{schmidt2017} and was an essential spring-board for further improvements.
    
    \item \textbf{Chemical element substitutions}. Using arguments of chemical similarity~\cite{glawe_optimal_2016} one can transmute chemical elements in a given crystal structure to obtain closely related compounds. This method was used with success in Ref.~\cite{wang2021predicting}, and led to the creation of the Wang-Botti-Marques (WBM) dataset that is now used extensively to benchmark uMLIPs.
    
    \item \textbf{Random search}. Crystal structures can be generated randomly, with tools such as PyxTal~\cite{fredericks2021pyxtal}, although the process usually adds chemical or structural constrains to increase the probability of yielding stable compounds. For example, one can impose charge neutrality, which can be easily computed from the stoichiometry and the oxidation states, or one can impose the so-called Pauling test for neutral materials which requires that positive ions have lower electronegativity than negative ones. This approach has been used, for example, in Ref.~\cite{2ddatabase}. This process can be seen as a very simplified method for crystal structure prediction~\cite{oganov_modern_2011}. Of course, more sophisticated methods, such as genetic algorithms~\cite{oganov_how_2011}, particle swarm methods~\cite{calypso}, the minima hopping method~\cite{goedecker_minima}, etc. can also be utilized.

    \item \textbf{Generative models}. Generative models for crystal structure prediction represent a paradigm shift from traditional approaches, as they learn the underlying distribution of thermodynamically stable crystal structures and sample directly from it, rather than modifying existing compounds. Such models---including variational autoencoders, generative adversarial networks, diffusion models, and flow-matching frameworks---are trained on crystal databases and can propose entirely new atomic arrangements and compositions that have never been synthesized or computed before. See Refs.~\cite{breuck_generative_2025,handokoArtificialIntelligenceGenerative2025} for a comprehensive review on the subject. We use our previously developed model, \textsc{Matra-Genoa}~\cite{breuck_matra_2025}, an autoregressive transformer architecture built on invertible tokenized representations of symmetrized crystals, including free coordinates. The model is computationally efficient, generates structures with correct symmetry, and produces compounds that cluster near the convex hull of thermodynamic stability. For a detailed description of its methodology and performance, we refer to the original publication~\cite{breuck_matra_2025}.
\end{itemize}

Spectacular advances in this field of uMLIPs have been made in the past couple of years, fueled by the developed of new models and by the availability of datasets like OMat24~\cite{omat24}, Materials Project~\cite{materialsproject}, and \alex{}. We have used two such uMLIPs for the preliminary relaxation of the crystal structure:
\begin{itemize}
    \item \textbf{\textsc{M3GNet}}~\cite{m3gnet}. This represents one of the first truly universal MLIPs based on graph neural networks, capable of accurately predicting energies, forces, and stresses for materials across the entire periodic table while maintaining exceptional computational efficiency that enables molecular dynamics simulations and structure optimizations at scales previously accessible only to classical force fields.
    \item \textbf{\textsc{Orb-v2}}~\cite{neumann_orb_2024}. In late 2024, the quality of uMLIPs improved drastically and far surpassed the performances of \textsc{M3GNet}. We chose to integrate \textsc{Orb-v2} model~\cite{neumann_orb_2024} as it presented the most favorable trade-off between accuracy and computational cost.
\end{itemize}
Over the past year, numerous new uMLIPs have been created, and some boast enhanced features~\cite{Matbench} compared to either \textsc{M3GNet} or \textsc{Orb-v2}. Consequently, it is probable that we will utilize different uMLIPs soon.

To predict the distance to the convex hull, from structures relaxed with the uMLIPs, we trained two different models:
\begin{itemize}
    \item \textbf{\textsc{FAENet}}. The Frame Averaging Equivariant Network is a graph neural network architecture for materials modeling that achieves $E(3)$ equivariance/invariance via data transformations rather than by building symmetry constraints into the model itself. The model trained used as input the \textsc{M3GNet} geometries and as output the DFT distance to the hull.
    \item \textbf{\textsc{ALIGNN}}. The Atomistic Line Graph Neural Network is a graph neural network model for materials science that represents atoms as graph nodes and explicitly includes both bonds and bond angles by constructing a line graph of the atomic structure. We trained two different models, the first used as input \textsc{M3GNet} geometries while the second used the DFT geometries. The output was the DFT distance to the convex hull.
\end{itemize}
Note that by using approximate geometries as inputs we could circumvent to a large extent the relatively large error of \textsc{M3GNet}. On the other hand the \textsc{Orb-v2} geometries are already so accurate that we found no advantage in using them for training. Details on the training procedure can be found in \Cref{sec:methods}.

Finally, in what concerns the DFT calculations, details are also presented in \Cref{sec:methods}.

\subsection{Benchmark of the workflow}

\begin{table*}[htb!]
\caption{Summary of the DFT validations on different sets, note the \ehull{} is in unit of meV/atom. To circumvent numerical issues, we define compounds on the hull as those that have \ehull$<1$~meV/atom.\label{tab:alexruns}}
\begin{tabular*}{\textwidth}{@{\extracolsep{\fill} }rrrrrrrr}
\toprule
Sets     & Total & \multicolumn{2}{c}{\ehull$<1$} &  \multicolumn{2}{c}{\ehull$<50$} & \multicolumn{2}{c}{\ehull$<100$} \\
\midrule 
\texttt{m3gnet/rng} & 29671 & 228 & (0.8\%) & 3868 & (13.0\%) & 10743 & (36.2\%) \\
\texttt{m3gnet/m3gnet} & 79078 & 1635 & (2.1\%) & 14447 & (18.3\%) & 36419 & (46.1\%) \\
\texttt{m3gnet/faenet} & 68734 & 3500 & (5.1\%) & 52116 & (75.8\%) & 64673 & (94.1\%) \\
\texttt{m3gnet/alignn} & 142328 & 39043 & (27.4\%) & 132200 & (92.9\%) & 139813 & (98.2\%) \\
\texttt{orb/alignn} & 345827 & 53385 & (15.4\%) & 331373 & (95.8\%) & 343242 & (99.3\%) \\
\bottomrule
\end{tabular*}
\end{table*}

\begin{figure}[htb]
    \centering
    \includegraphics[width=8cm]{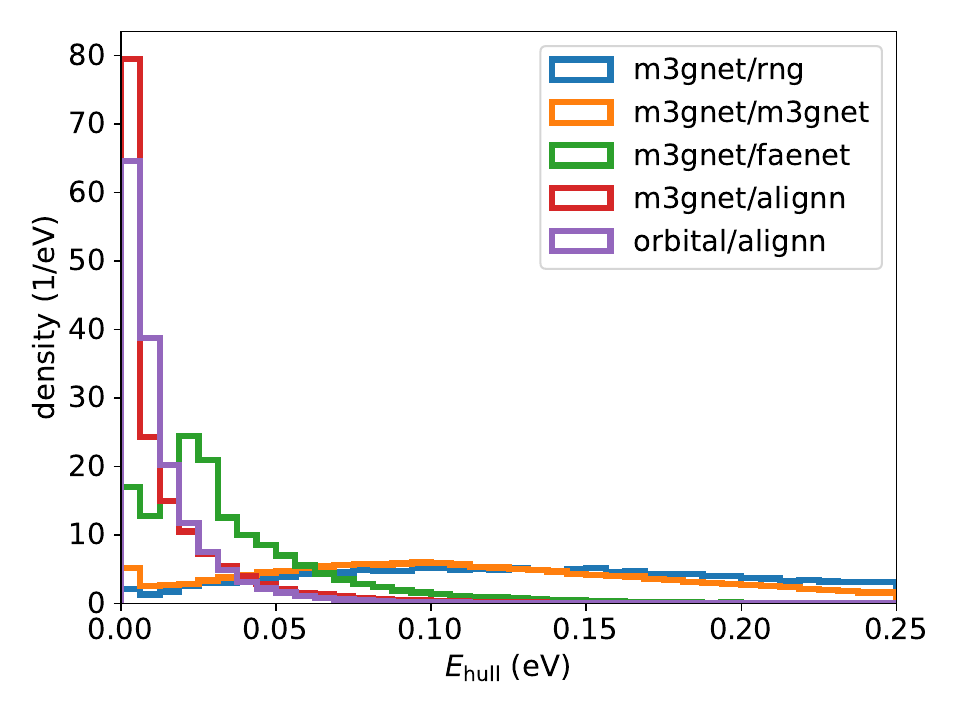}
    \caption{Histograms of the distances to the convex hull of the different datasets described in the text. The histograms are normalized for easier comparison. }
    \label{fig:alexruns}
\end{figure}

In the following, we show how the efficiency of our workflow to discover compounds close to the convex hull of stability improved. In order to provide a fair comparison we prepared a control dataset by performing chemical substitutions on compounds with $E_\text{Hull} < 100$~meV/atom that were present in \alex{}, following the recipe of Ref.~\cite{wang2021predicting}. In total we generated more than 70 million crystal structures that were then optimized with the \textsc{M3GNet}~\cite{m3gnet} model.
From these a random sample of 42 thousands structures was relaxed with DFT. We label this control dataset \texttt{m3gnet/rng} in \Cref{tab:alexruns} and \Cref{fig:alexruns}. We take the opportunity to note that these results were obtained with respect to the updated (self-consistent) convex hull.

In this control run, we observe a smooth distribution of distances to the hull, with a maximum centered at around 100~meV/atom, and with a fat tail slowly decaying to energies above 500~meV/atom.
We observe a success rate of 0.8\% in finding compounds on the convex hull and 36.2\% in identifying compounds within 100~meV/atom from the hull.
These values are lower than those reported in Ref.~\cite{wang2021predicting}, which can be attributed to two main factors: first, our substitutions were performed not only on stable compounds (as in Ref.~\cite{wang2021predicting}) but also on compounds up to 100~meV/atom above the convex hull; second, the convex hull has become significantly more complete since 2021.
Nonetheless, these results already outperform random sampling of the entire materials space by a factor of ten~\cite{schmidt2017}.

To benchmark the importance in the estimation of the distance to the hull, we prepared three further datasets using different strategies.
\begin{enumerate}
    \item We used the \textsc{M3GNet} energy to compute directly the distance to the DFT convex hull. We then selected the 88 thousands structures compounds predicted to have the smallest distance to the hull for further DFT relaxation. This dataset is labeled \texttt{m3gnet/m3gnet} in \Cref{tab:alexruns} and \Cref{fig:alexruns}.
    \item We used the \textsc{FAENet} model to predict the distance to the convex hull and selected 69 thousands compounds for further DFT relaxation that were closest to the hull. This dataset is labeled \texttt{m3gnet/faenet} in \Cref{tab:alexruns} and \Cref{fig:alexruns}.
    \item We used the \textsc{ALIGNN} model to predict the distance to the convex hull and selected 143 thousands compounds for further DFT relaxation that were closest to the hull. This dataset is labeled \texttt{m3gnet/alignn} in \Cref{tab:alexruns} and \Cref{fig:alexruns}.
\end{enumerate}

The analysis of the DFT energies showed clearly that distance to the hull obtained directly from the \textsc{M3GNet} was not a good estimator of the final stability.
We see already a clear improvement when using the \textsc{FAENet} with 94.1\% of the dataset validated to be below 100~meV/atom form the DFT convex hull. However, the most remarkable result comes form using the \textsc{ALIGNN} model, with 27.4\% of the compounds being on the DFT convex hull of stability and 98.2\% below 100~meV/atom.
The distributions of \texttt{m3gnet/faenet} and \texttt{m3gnet/alignn} sets are strongly peaked at zero (or close to it) and decay rapidly with essentially no compound found with distances to the hull larger than around 150~meV/atom.
This approach allows us to circumvent to a large extent the errors stemming from the uMLIP relaxation and energy prediction.
The superior performance of these models can be attributed to their focus on local energy minima rather than the entire potential energy surface, effectively reducing the input space and improving prediction accuracy.

In \Cref{tab:alexruns} and \Cref{fig:alexruns} we also present a dataset composed of structures relaxed with \textsc{Orb-v2} which were predicted to be close to the convex hull of stability by an \textsc{ALIGNN} model.
Due to the high quality of the \textsc{Orb-v2}, this \textsc{ALIGNN} model was trained to predict the DFT distance to the hull from the DFT geometry.
While \textsc{Orb-v2} produces structures significantly closer to the PBE equilibrium geometry than \textsc{M3GNet}, the resulting distributions of distances to the convex hull are remarkably similar between the two. This can be attributed to the ability of the structure-property models to learn the mapping between \textsc{M3GNet}-relaxed geometries and PBE-level distances to the hull, even in the presence of structural errors introduced by the less accurate force field. In fact, achieving high accuracy in a structure-property model is much simpler than training an accurate uMLIP, as the former needs only to capture the configuration space of mechanically stable structures, which is substantially smaller and better-behaved than the full configuration space sampled during dynamics. Regardless, employing \textsc{Orb-v2} geometries as input to DFT yields a practical advantage, as the higher quality of the initial geometry leads to faster convergence in the DFT geometry relaxation.

In summary, our current optimized workflow uses i)~\textsc{Matra-Genoa}~\cite{breuck_matra_2025} to generate candidate compounds.
We do not restrict in any way the chemical composition, the space group, the number of crystal structures per stoichiometry, etc.;
ii)~geometry pre-relaxation using \textsc{Orb-v2};
iii)~estimation of the distance to the hull using \textsc{ALIGNN}; and finally iv)~DFT relaxation.
That last step is by far the most computationally intensive, being several orders of magnitude slower than any machine-learning-based step.

\subsection{Data}

\begin{figure}[ht]
    \centering
    \includegraphics[width=\linewidth]{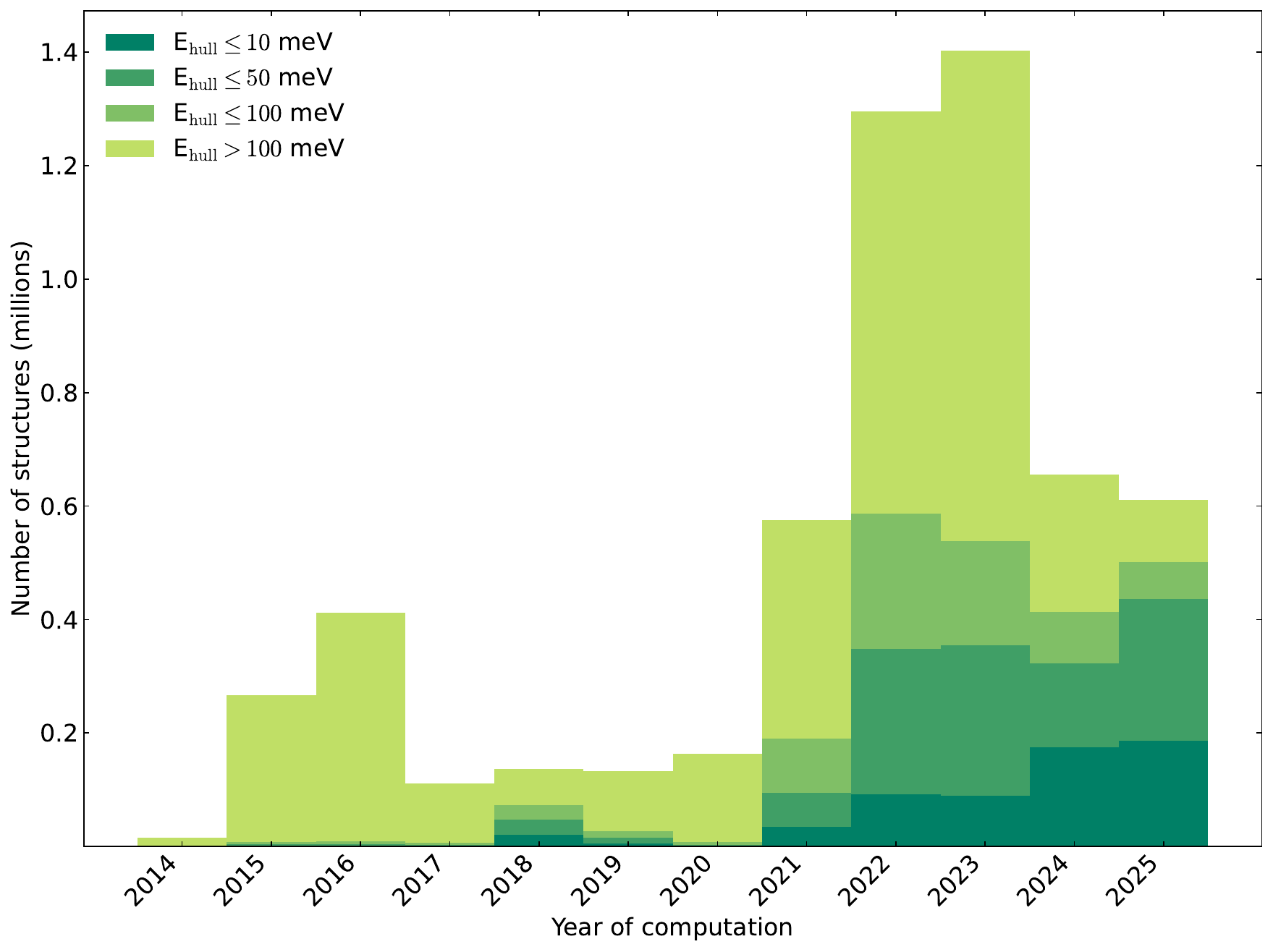}
    \caption{Evolution of the number of structures introduced to the database through time, in relation to their distance to the hull.}
    \label{fig:history}
\end{figure}

In \Cref{fig:history}, we illustrate the progress of the \alex{} database from its inception to July 2025 as an example of the broader advancements in machine learning for materials discovery. More importantly than the growth in the total number of entries, the percentage of compounds added to the database close to the convex hull of thermodynamic stability increased considerably over the years. The evolution can be seen through key milestones: from the exploration of the full compound space of perovskites and other families with small, symmetric, unit cells in 2017~\cite{schmidt2017}, to the development of prototype-specific models in 2018~\cite{jonathan2018}, to the introduction of crystal graph networks trained on millions of data points~\cite{cgat}, and finally to the current workflow incorporating generative models, as described above.

\begin{figure*}[t]
    \centering
    \includegraphics[width=0.9\textwidth]{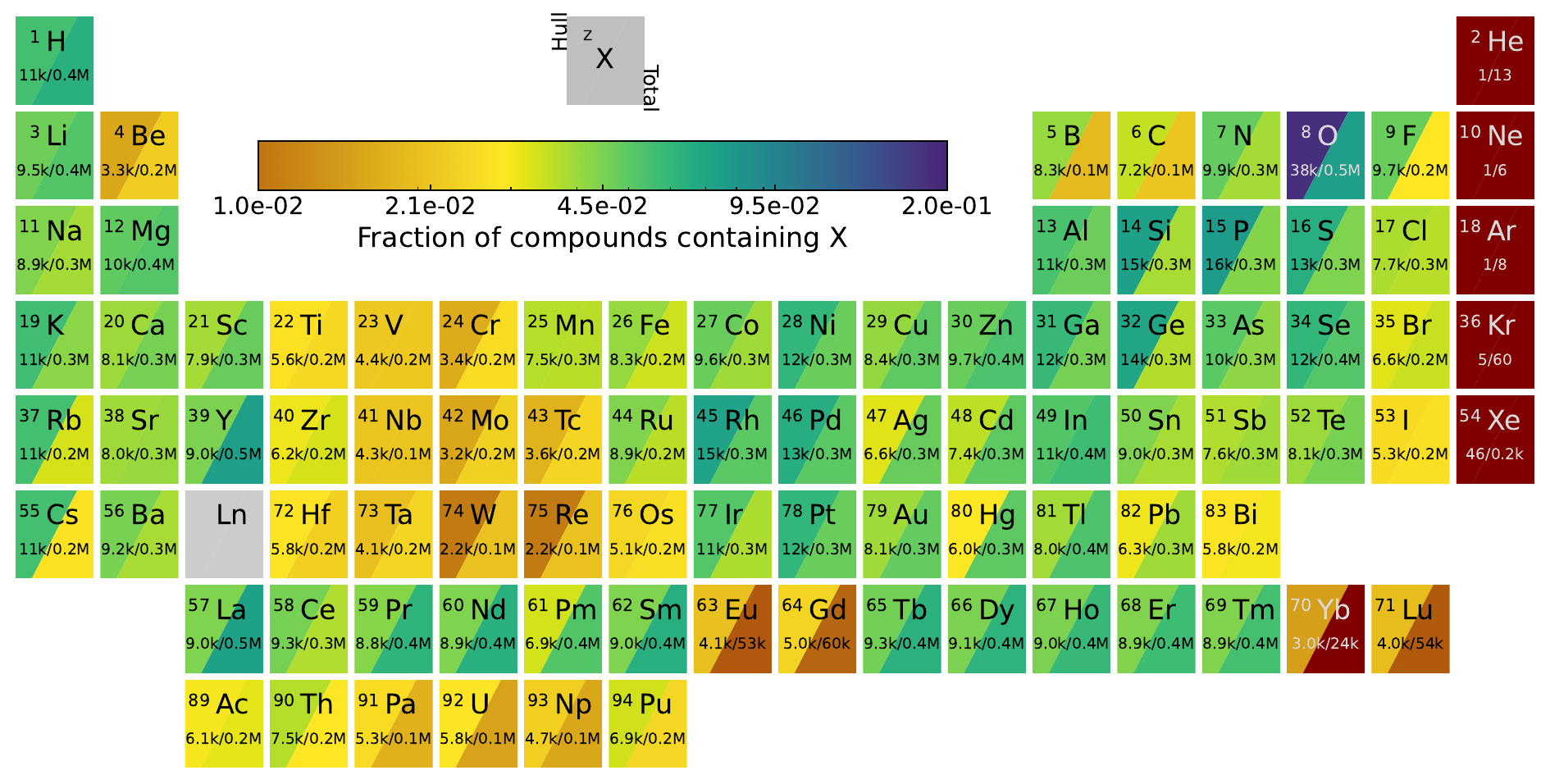}
    \caption{Elemental distribution within the database. Each cell of the periodic table indicates the number of materials on the convex hull (upper left) and in the entire database (lower right) containing a given element.}
    \label{fig:hull_fraction}
\end{figure*}

In \Cref{fig:hull_fraction} we compare the elemental distributions of the compounds in the convex hull of thermodynamic stability compared to the whole \alex{}. At a first sight, the distribution in the two datasets is rather consistent. By far, the most represented chemical element is oxygen. In our opinion this is due to two factors: (i)~the large chemical reactivity of \ce{O2}, easily forming compounds; and (ii)~the large number of experimentally known oxides in ICSD~\cite{zagorac_recent_2019}, that was used to seed materials databases, creating a bias that survived until today. Other non-metals and semi-metals, such as Si, Ge, P, S, or F are also well represented both in the hull and in \alex{}. Unsurprisingly, the rare gases, with their inability to form compounds, are the least common. In what concerns metals, alkali and alkali-earth metals (with the exception of Be) and late d-metals are all rather common. However, this is not the case with metals from groups IV-VIII that have more difficulty in forming chemical compounds. 
The lanthanide series presents an intriguing scenario because the occurrences of these chemical elements tend to remain relatively stable across the board. However, exceptions are noted with europium (Eu), gadolinium (Gd), ytterbium (Yb), and lutetium (Lu). This anomaly can largely be attributed to a specific artefact inherent in our computational calculations. Namely, the PAW setups employed by \textsc{Vasp}~\cite{paw2} for these four elements lead to complications that hinder the convergence of compounds containing them.

We note that systematic errors in the workflow used to generate and filter the compounds can also have an influence in the overall distribution of chemical elements in \alex{}. For example, we observe that the substitution process introduces a significant bias towards oxides, with 16\% of the substituted compounds containing oxygen. Hydrogen is also an interesting case, with the initial substitutions resulting in 22\% hydrogen-containing compounds, while the \textsc{M3GNet} set amplifies this number to 69\%. In contrast, \textsc{FAENet} and \textsc{ALIGNN} models predict less (meta-)stable compounds containing hydrogen, at 14\% and 4\%, respectively. Therefore, care has to be taken when interpreting elemental distributions such as the ones in \Cref{fig:hull_fraction}.

\begin{figure}[htb]
    \centering
    \includegraphics[width=\columnwidth]{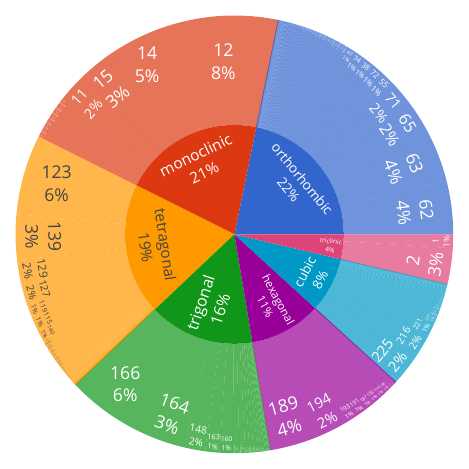}
    \caption{Space group distribution for of structures on the convex hull.}
    \label{fig:spg}
\end{figure}

In \Cref{fig:spg} we depict the distribution of the compounds on the convex hull of thermodynamic stability with respect to their space groups.
At a first sight we see that all crystal systems are present in the convex hull, although in percentages that differ from assuming an equal distribution of all space groups.
In particular we find less cubic ($36/230=15.7\%$ space groups) and tetragonal ($68/230=29.6\%$) compounds and more triclinic ($2/230=0.9\%$) and monoclinic ($13/230=5.7\%$) materials than what could be expected from a uniform distribution.
Note that part of the excess of low-symmetry crystal structures can be explained by imperfect geometry relaxations.
We would like to stress that many experimental materials exhibit often higher symmetry on average.
Recognizing that our database comprises stoichiometric, ordered compounds, this apparent discrepancy can be attributed to several distinct mechanisms. Compositional disorder, for instance, can average over symmetry-inequivalent Wyckoff positions, effectively raising the observed symmetry relative to that of the ordered ground state. Additionally, anharmonic effects and finite-temperature contributions may selectively stabilize higher-symmetry phases with respect to their zero-temperature harmonic counterparts.

While natural occurring minerals have been found in all space groups, we witness that some space groups are significantly more populated than others. For example, in the cubic system most compounds crystallize in groups \#225, \#216, \#221, and \#227, while no compound can be found in space groups \#207, \#208, \#210, \#211, or \#222. For hexagonal compounds, space groups \#189 and \#194 are extremely common, while we only find one compound in group \#196. In fact, this situation is not unique to stoichiometric inorganic compounds, but it is also found in organic crystals~\cite{wilson_space_1988,wilson_space_1990} or in proteins~\cite{wukovitz_why_1995}.

\begin{figure*}[htb]
    \centering
    \includegraphics[width=0.9\textwidth]{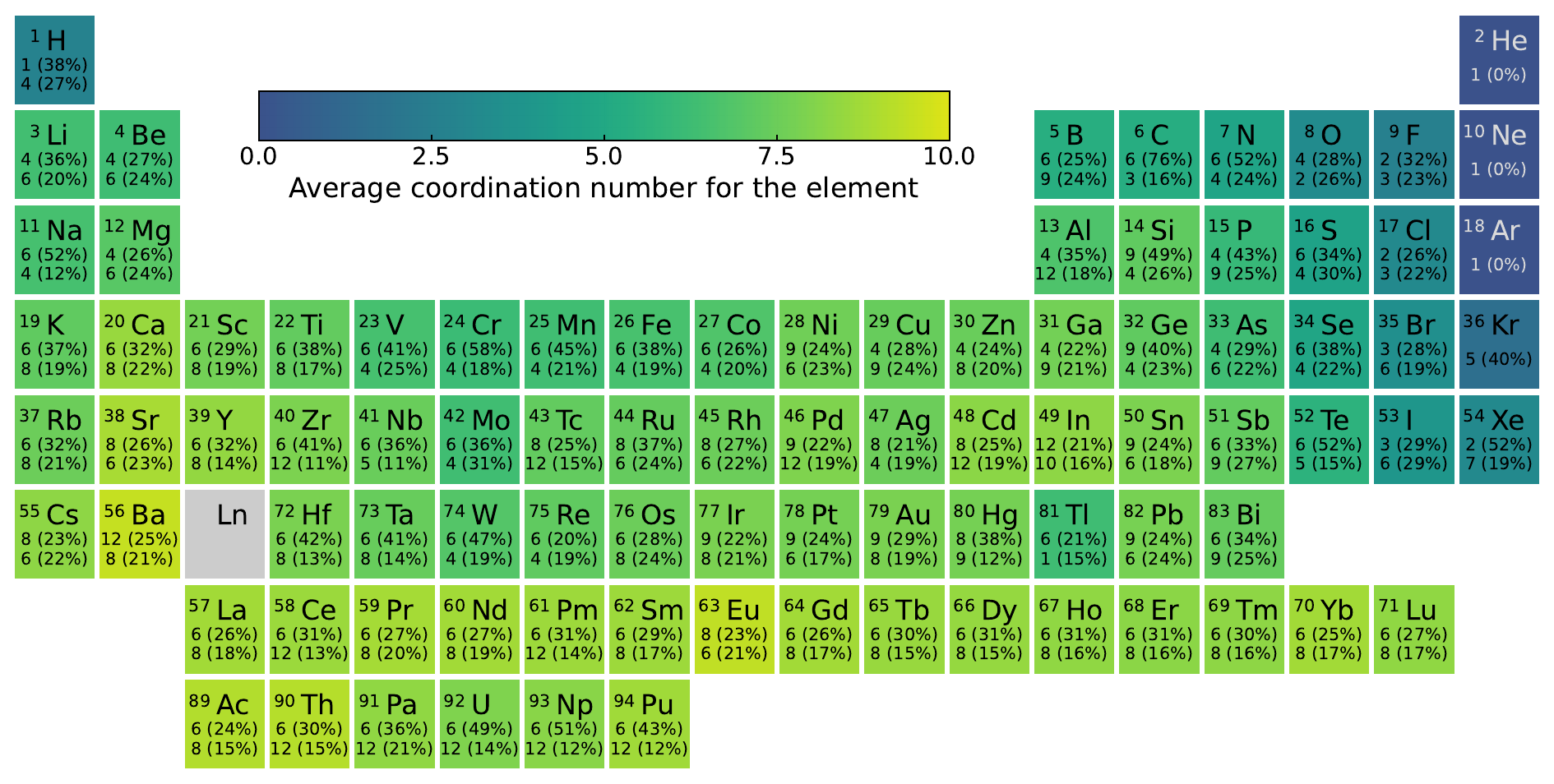}
    \caption{Average and most frequent coordination number of elements in structures on the convex hull. The background color represents the average coordination number, while the overlaid numbers indicate the most frequent coordination value (with the corresponding occurrence percentage shown in parentheses).}
    \label{fig:coord_period}
\end{figure*}

Beyond the global symmetry of crystal structures, local atomic coordination environments are fundamental to understanding material properties, as they directly influence electronic structure, bonding character, and physical behavior. To characterize these local environments across our database, we computed coordination numbers for each crystallographically distinct atomic site. Coordination numbers were determined using the \texttt{CrystalNN} method available in \texttt{pymatgen} as it has been shown to match human judgement of coordination well\cite{pan_benchmarking_2021,pymatgen}.
\Cref{fig:coord_period} presents the most common coordination numbers (CN) for each element across the periodic table, revealing well-established chemical trends consistent with fundamental bonding principles. For instance, many transition metals, such as titanium, vanadium, chromium and iron predominantly adopts the expected octahedral coordination (CN = 6), while halogens favor low coordinations.
The data also illustrate the systematic tendency for larger ionic radii to accommodate higher coordination numbers, as can be seen, for example, by the progressive increase observed across the alkaline earth metals.

Interestingly some unexpected distributions also appear.
For example, while tetrahedral silicon (CN = 4) is abundant (26\%), a large portion of silicon structures (49\%) shows a coordination of 9. Similarly, the coordination of aluminium does not exhibit the octahedral environment as prominently as one might expect (as found in materials such as $\alpha$-alumina), but instead shows tetrahedral (CN = 4) and CN = 12 coordination. This distribution reveals that our database contains a substantial fraction of structurally unconventional materials, underscoring the rich diversity of coordination environments that can give rise to exceptional properties.
As an example of these unconventional configurations we can mention the high occurrence of six-fold coordinated carbon and nitrogen that are characteristic signatures of, for example, antiperovskite compounds~\cite{krivovichev_structural_2024}.
Substitutional disorder has recently emerged as a major concern in AI-driven materials databases~\cite{disorder2}, including \textsc{GNoME}~\cite{gnome}. To assess this effect within the \alex{} database, we applied a recently developed machine-learning model for classifying ordered and disordered crystal structures based solely on composition and experimental training data~\cite{disorder} to all compounds on the convex hull. As summarized in \Cref{tab:disorder}, \alex{} exhibits only slightly higher predicted disorder rates than the Materials Project or \textsc{ICSD}, and significantly lower rates than the \textsc{GNoME} dataset. Because the GNoME dataset predominantly contains quaternary and quinary compounds, which are intrinsically more prone to disorder, we further compared predicted disorder rates for ternary compounds in both datasets. Among the 86,000 ternaries in \alex{}, 35–39\% are predicted to be disordered, compared to 59–65\% in 63000 ternaries in the \textsc{GNoME} database. These results indicate that \textsc{GNoME} exhibits a strong bias toward disordered structures, largely independent of composition complexity. According to these predictions, the \alex{} convex hull already includes a larger number of new ordered crystal structures than the \textsc{GNoME} database.

\begin{table}[ht]
\centering
\caption{Estimated fraction of substitutionally disordered compounds across major crystal databases. Results for Materials Project, GNoME, and ICSD are taken from Ref.~\cite{disorder}. Data for Alexandria was predicted using the RNN model~\cite{disorder} with decision boundaries at 50\% and 70\% probability as in Ref.~\cite{disorder}. The value from the experimental ICSD is taken as a baseline to compare the other values.}
\label{tab:disorder}
\begin{tabular}{lcc}
\toprule
\textbf{Database} & \textbf{Estimated Disorder (\%)} \\
\midrule
Materials Project (MP) & 25--31  \\
GNoME & 80--84  \\
\textbf{Alexandria} & \textbf{37--43}  \\
ICSD & 35\\
\midrule
\bottomrule
\end{tabular}
\end{table}

\begin{figure*}[ht]
    \centering
    \includegraphics[width=0.9\textwidth]{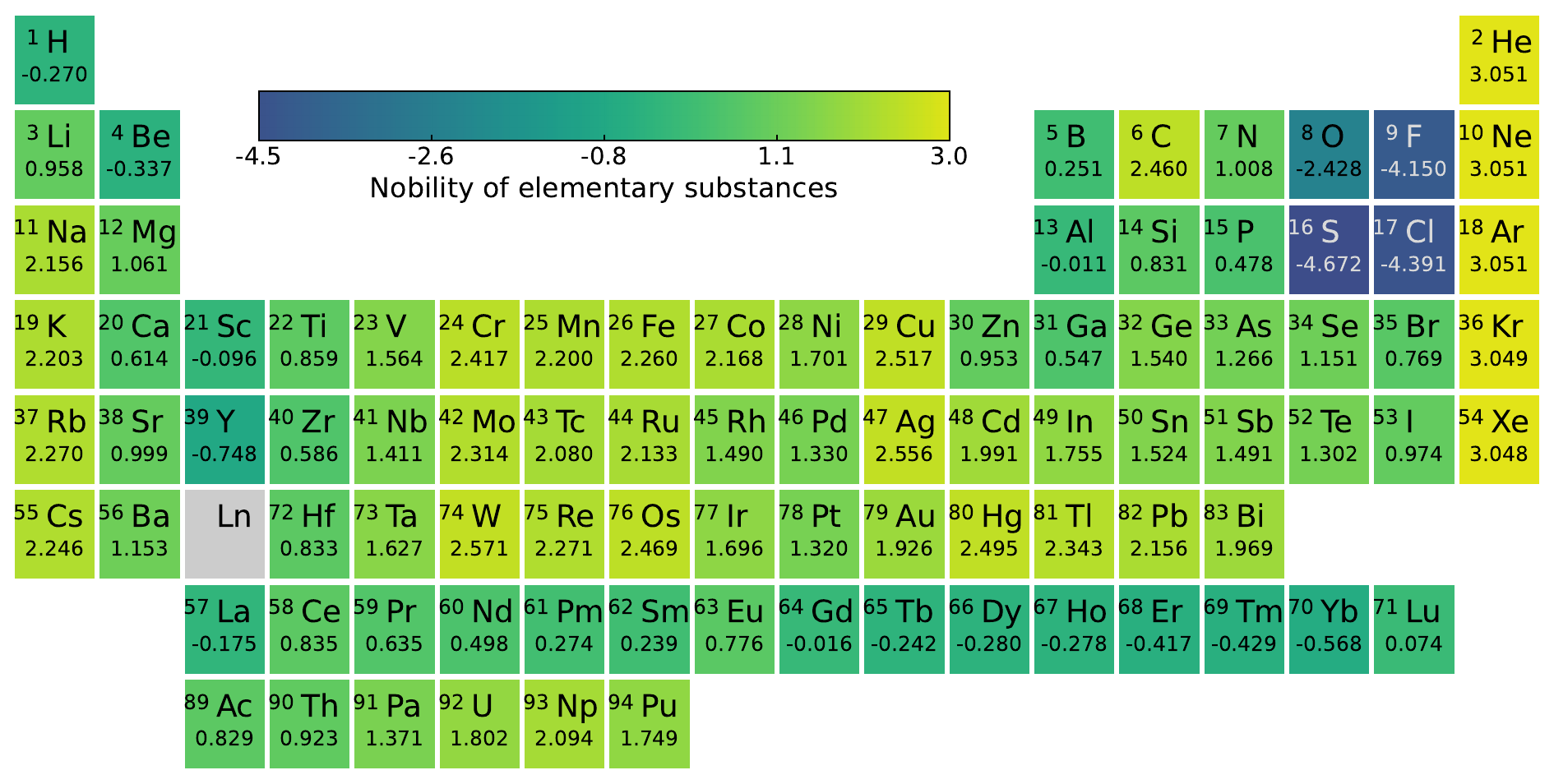}
    \caption{Nobility index of elementary substances on the convex-hull of \alex{}. High nobility indicates that the material is in a two-phase equilibrium with many other materials of the convex hull, while low nobility indicates that the material reacts with many materials.}
    \label{fig:nobility_period}
\end{figure*}

Beyond intrinsic material properties, the scale of our database also allows for the investigation of emergent ``collective'' properties that arise from inter-material relationships rather than intrinsic characteristics of individual compounds. The connectivity patterns within the convex hull reveal how compounds organize into networks of phase equilibria, providing insights into which materials play central roles in synthesis pathways and decomposition reactions.
One such property is nobility, a graph-theoretic measure introduced by Wolverton \textit{et al.}~\cite{wolverton2020network} that quantifies the degree of connectivity of a material within the convex hull of thermodynamic stability. Specifically, nobility is defined as the fraction of all possible edges (representing two-phase equilibria) in which a given compound participates as a stable phase. Materials with high nobility values are in phase equilibria with many others, functioning as central hubs of the stability network. These highly connected compounds frequently emerge as decomposition products or stable intermediate phases during synthesis and processing. Conversely, compounds with low nobility occupy peripheral positions in the network, participating in fewer phase equilibria and often representing more reactive compositions. This metric thus provides a quantitative framework for understanding which materials play important roles in organizing chemical phase space.

Earlier investigations into the concept of nobility, as documented in~\cite{wolverton2020network}, were restricted by the limited scope and chemical variation present in the available materials databases. Our recently developed and considerably enlarged convex hull, which incorporates nearly eight times the number of compounds and 27 times the network edges compared to the one detailed in Ref.~\citenum{wolverton2020network}, facilitates a thorough and systematic inquiry into the progression of nobility values when computed over a markedly broader and more chemically diverse array of materials. This enhancement also provides us with the opportunity to discern whether novel patterns become apparent at this expanded scale.

An intriguing consequence of database scale is the sub-linear growth of the average node degree (number of neighbors) within the stability network. Our expanded network exhibits an average of $12.5\times10^3$ neighbors per node---only approximately three times larger than the original dataset~\cite{wolverton2020network}, despite the substantially increased number of materials. This sub-linear scaling suggests a form of saturation in the local connectivity of nodes, where further expansion of the convex hull may asymptotically approach a limiting average degree. Regarding the nobility values themselves, the enlarged dataset offers both opportunities and challenges: while statistical biases diminish with increased data, any systematic deficiencies in the concept of nobility become more pronounced. 

Looking at \Cref{fig:nobility_period} we observe that the most salient features of nobility are preserved with the larger convex hull. Expected trends remain evident, including low nobility for highly electronegative elements (\ce{O} and \ce{F}), maximum nobility for noble gases, and systematic gradients across the transition metal block. However, comparison with previous result reveals scale-dependent artifacts. Extremely low-nobility species such as \ce{F} and \ce{Cl} exhibit increasingly negative values with a larger hull, while the counterintuitive observation that Au has lower nobility than Fe persists. We conclude therefore, that these deficiencies are not a consequence of the small size of the hull in Ref.~\citenum{wolverton2020network} but are due to the definition of nobility.
We believe that deficiencies might be resolved by introducing a refined nobility metric that consistently weights network edges to achieve size invariance, thereby providing a more robust measure of thermodynamic centrality across databases of varying scale.

\subsection{Relaxation paths}

\begin{figure}
    \centering
    \includegraphics[width=\linewidth]{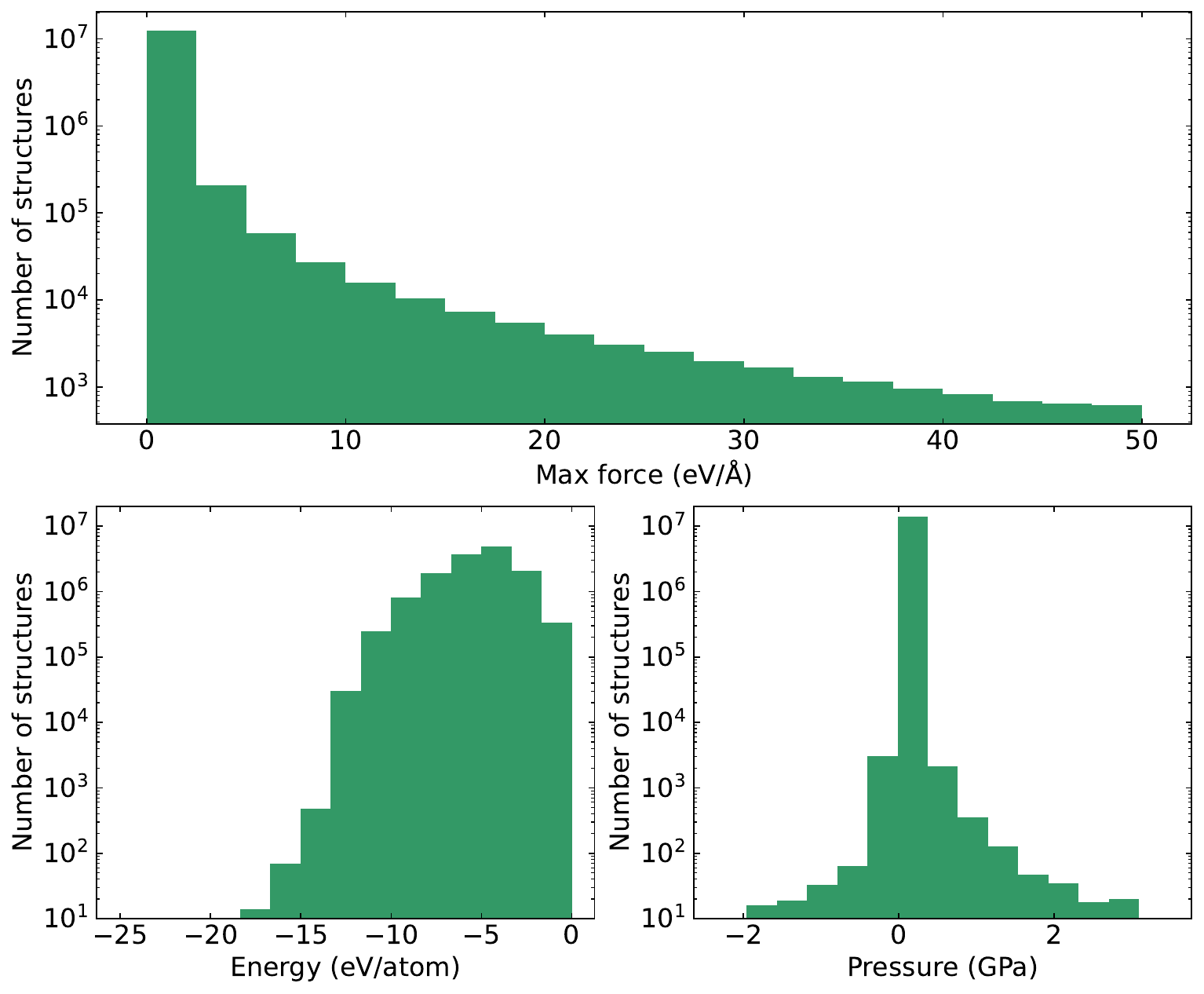}
    \caption{Distribution of maximum forces, energies and pressure across \texttt{sAlex25} (subsampled dataset).}
    \label{fig:subalex_histograms}
\end{figure}

A primary application of the \alex{} database is for training of uMLIPs~\cite{Matbench}. Such models provide accurate predictions of energies, forces and stresses across the complete potential energy surface, often with accuracy comparable to DFT calculations~\cite{bhatia_mace4ir_2025,loew_universal_2025,benedini_universal_2025,loew_pressure_2025}, but at a small fraction of the computational costs~\cite{peng_lambench_2025}. Typically, state-of-the-art uMLIPs are pre-trained in the OMAT24 dataset~\cite{omat24} and then fine-tuned in geometry relaxation trajectories from the Materials Project~\cite{materialsproject} and \alex{}.

In their 2024 work, Barroso-Luque \textit{et al.} proposed a procedure to filter these steps based on energy differences along the relaxation path and published the dataset now known as subsampled \alex{} (sAlex)~\cite{omat24}. Alongside with this update of \alex{}, we provide a new version of the subsampled database following the same procedure. Specifically, from each geometry optimization path we retain only the first and last steps, and any other structures that differ in energy by more than 10~meV/atom. Furthermore, all materials present in the Wang-Botti-Marques dataset~\cite{wang2021predicting} (commonly used as a benchmark) as well as all outliers with positive total energy or abnormally large forces and stresses were removed. This dataset, that we name \texttt{sAlex25}, contains around 14 million out-of-equilibrium structures along structure relaxation trajectories complete with energy, forces and stress.

To illustrate the diversity of configurations represented in the \texttt{sAlex25}, we show in \Cref{fig:subalex_histograms} the distribution of forces, energies and pressures across the dataset.
The dataset is heavily biased toward dynamically stable structures, as is to be expected from the generation process. This makes it ideal for fine-tuning uMLIPs for accurate prediction of properties close to the minimum energy structure, such as equilibrium geometries, phonons, mechanical and thermal properties, etc.

\subsection{Graph Atomic Cluster Expansion}

To estimate the influence of the extended \texttt{sAlex25} on uMLIPs training, we use it in combination with Materials Project dataset to fine-tune a \textsc{GRACE-2L-OMAT-L} model pretrained on the OMat24 dataset, following the procedure and naming convention as described in \cite{lysogorskiyGraphAtomicCluster2025}. The resulting \textsc{GRACE-2L-OAM-L} achieves a F1 score of 0.895 and a MAE of 21.2~meV/atom on the WBM benchmark~\cite{wang2021predicting,Matbench}. 
These improved metrics suggest that \texttt{sAlex25} offers a slight enhancement of the uMLIP description of equilibrium structures. Given that \texttt{sAlex25} is considerably larger than its predecessor, this suggests that the training data near dynamical equilibrium may already be saturated, with uMLIPs unable to significantly benefit from additional data in this regime.

\section{Conclusions}

This work presents a major expansion of the \alex{} database, now comprising 5.8 million DFT-calculated structures with 175 thousand thermodynamically stable compounds on the convex hull, exhibiting predicted rates of structural disorder consistent with experimental inorganic crystal structure databases. More significantly, we have developed a highly efficient multi-stage discovery workflow that combines generative models, universal machine learning interatomic potentials, and specialized graph neural networks to achieve an unprecedented 99\% success rate in identifying compounds within 100~meV/atom of thermodynamic stability.

Our systematic benchmarking demonstrates that the key innovation lies not in using more powerful individual models, but in the strategic orchestration of complementary machine-learning approaches: \textsc{Matra-Genoa} for targeted structure generation, \textsc{Orb-v2} for accurate preliminary relaxations, and \textsc{ALIGNN} for precise energy-to-hull prediction from approximate geometries. This workflow eliminates the accuracy bottleneck of preliminary relaxations through specialized prediction models, enabling us to generate 119 million candidate structures and add 1.3 million DFT-validated compounds to the database, including 74 thousand new thermodynamically stable materials. The computational efficiency gains are substantial: our multi-stage filtering reduces the number of required DFT calculations by over two orders of magnitude compared to random sampling approaches.

The updated subsampled \alex{} dataset (sAlex25) provides 14 million out-of-equilibrium structures with forces and stresses along relaxation trajectories, specifically curated for training universal machine learning interatomic potentials. We demonstrate the dataset's utility by fine-tuning a GRACE model, achieving state-of-the-art performance on the WBM benchmark. This resource addresses a critical need in the community for diverse, high-quality training data near equilibrium geometries.

Beyond database expansion, we have characterized fundamental patterns that emerge at this unprecedented scale. Our analysis reveals systematic trends in space group distributions, element-specific coordination preferences, and phase stability networks. Interestingly, the sub-linear scaling of convex hull connectivity suggests approaching saturation in local phase relationships, providing insights into the organization of thermodynamic stability across chemical space. The low predicted disorder rates (37--43\%) in \alex{}, comparable to experimental databases and significantly lower than other recent AI-generated datasets, underscore the quality and reliability of our discovered materials.

We emphasize our commitment to open science: all data, models, and computational workflows are released under permissive Creative Commons licenses to enable unrestricted use for research, model training, and further development. Unlike other recent large-scale materials discovery efforts, we make our entire generated dataset (not just the convex hull) publicly available, providing the community with unprecedented resources for developing and benchmarking improved discovery methodologies. This transparency enables reproducibility and invites critical evaluation, fostering a truly collaborative ecosystem for materials discovery.

The \alex{} database now stands as the most comprehensive open foundation for data-driven materials discovery, combining scale with quality and validated computational workflows with practical demonstrations. Looking forward, we anticipate that continued improvements in generative models and force fields trained on \alex{} will create a positive feedback loop, further accelerating the discovery of functional materials for energy, catalysis, and electronics applications. We invite the community to leverage this resource for training next-generation models, discovering novel materials, and developing innovative methodologies that will transform computational materials design.

\section{Methods}
\label{sec:methods}

\subsection{DFT}
\label{sec:dft}
We performed geometry optimisations and total energy calculations using the \textsc{Vasp} code~\cite{vasp1,vasp2}. All parameters, including pseudopotentials, were set to ensure compatibility with the data available in the Materials Project database~\cite{materialsproject}.  To sample the Brillouin zones we used uniform $\Gamma$-centred k-point grids with a density of 1000 k-points per reciprocal atom. Spin-polarised calculations were started from a ferromagnetic configuration. We used the projector augmented wave (PAW) setup~\cite{paw,paw2} within \textsc{Vasp} version 5.2, applying a cutoff of 520~eV for all materials. We set the convergence criteria of the forces to be less than 0.005~eV/\AA. We used the Perdew-Burke-Ernzerhof (PBE) exchange correlation functional~\cite{PBE} with on-site corrections for oxides and fluorides containing Co, Cr, Fe, Mn, Mo, Ni, V or W for all 1D, 2D and 3D compounds. The U corrections for the d-states were 3.32, 3.7, 5.3, 3.9, 4.38, 6.2, 3.25 and 6.2 eV respectively. Some calculations did not converge due to various problems, and these entries were subsequently removed from the database.

\subsection{\textsc{FAENet}}
To limit the maximum memory usage we removed structures with more than 30 atoms from the training and validation set, resulting in 3.5 million training and 440 thousand validation systems.
We used the \textsc{FAENet} implementation in the \textsc{Intel MatSciML} framework~\cite{lee2023matsciml} running on 32 compute nodes with one P100 each, using a total batch size of 16x32, resulting in 512 samples per batch.
The structures were encoded using a cutoff of 6~\AA\, and a maximum of 40 neighbors per node using 5 interaction layers, 104 Gaussian basis functions for radial basis expansion, 128 channels for the positional embeddings, 480 filters in the interaction blocks and hidden dimension of 512.
The final node embeddings were reduced using a mean pooling strategy.
The output network used 3 hidden layers with size 512, SiLU activations~\cite{silu}, skip connections and layer normalization to stabilize training.
For training we used AdamW~\cite{adamw} with a learning rate of 0.0005 and a cosine annealing learning rate scheduler configured with a minimum learning rate of 0.00001 and a cycle length of 130 epochs.

\subsection{\textsc{ALIGNN}}
The \textsc{ALIGNN} model was trained in a dataset of 4.4 million datapoints with a range from 0 to 8~eV, and a mean absolute deviation from the mean of 0.39~eV.
From this dataset we split 300 thousand entries for validation and 300 thousand for testing, while the remaining 3.8 million entries were used for training.
We performed 200 epochs, and the best model according to the validation error was selected for the following.
The MAE of the \textsc{ALIGNN} model that predicted the distance to the hull from \textsc{M3GNet} relaxed structures was 33~meV/atom, considerably worse than the 16~meV/atom of the same model trained with relaxed PBE structures.
We believe that the increase of error comes from some structures for which \textsc{M3GNet} has blind spots and yields unreasonable geometries.

\subsection{GRACE}
The Graph Atomic Cluster Expansion (\textsc{GRACE}) \cite{bochkarevGraphAtomicCluster2024} provides a rigorous mathematical foundation for machine learning interatomic potentials, unifying both local and message-passing graph neural network (GNN) architectures in a single framework. \textsc{GRACE} generalizes the Atomic Cluster Expansion (ACE) \cite{drautzAtomicClusterExpansion2019}, which builds on a complete basis for local, star graphs, by introducing a complete set of tree graph cluster basis functions. This generalization enables the systematic inclusion of local and semi-local interactions. 
Tensor decomposition of expansion coefficients facilitates chemical embedding, enabling a single \textsc{GRACE} parameterization to describe all chemical interactions. Recursive evaluation further derives equivariant message passing GNNs from \textsc{GRACE} and places \textsc{GRACE} models on the accuracy-efficiency Pareto front of current uMLIPs \cite{lysogorskiyGraphAtomicCluster2025}. We employed a training procedure as described in  \cite{lysogorskiyGraphAtomicCluster2025}. Details of \textsc{GRACE} and its relation to other methods can be found in \cite{bochkarevGraphAtomicCluster2024}.

\section{Data availability}
\label{sec:data_ava}
The \alex{} can be accessed and/or downloaded from \url{https://alexandria.icams.rub.de/} under the terms of the \href{https://creativecommons.org/licenses/by/4.0/}{Creative Commons Attribution 4.0 License}.  

\section{Code availability}
\label{sec:code_ava}
All code and models developed in this work will be freely available upon acceptance at  \url{https://github.com/hyllios/utils/tree/main/}. The \textsc{GRACE} software and parametrizations are available at \url{https://gracemaker.readthedocs.io}. Pymatviz~\cite{pymatviz} was used to create the \Cref{fig:spg}.

\section{Acknowledgements}

A.L., S.B., and M.A.L.M. acknowledge funding from the Horizon Europe MSCA Doctoral network grant n.101073486, EUSpecLab, funded by the European Union. M.A.L.M. was supported by a collaboration between the Kavli Foundation, Klaus Tschira Stiftung, and Kevin Wells, as part of the SuperC collaboration, and by the Simons Foundation through the Collaboration on New Frontiers in Superconductivity (Grant No. SFI-MPS-NFS-00006741-10). S.B. acknowledge funding from the Volkswagen Stiftung (Momentum) through the project ‘‘dandelion''. The authors gratefully acknowledge the computing time made available to them on the high-performance computers Noctua and Otus at the NHR Center Paderborn Center for Parallel Computing (PC2). This center is jointly supported by the Federal Ministry of Research, Technology and Space and the state governments participating in the National High-Performance Computing (NHR) joint funding program (\url{www.nhr-verein.de/en/our-partners}). We gratefully acknowledge the Gauss Centre for Supercomputing e.V. (www.gauss-centre.eu) for funding this project by providing computing time on the GCS Supercomputer SUPERMUC-NG at Leibniz Supercomputing Centre (\url{https://www.lrz.de}). We also thank the Pittsburgh Supercomputer Center (Bridges2) and San Diego Supercomputer Center (Expanse) through allocation DMR140031 from the Advanced Cyberinfrastructure Coordination Ecosystem: Services \& Support (ACCESS) program, which is supported by National Science Foundation grants No. 2138259, No. 2138286, No. 2138307, No. 137603, and No. 2138296. A.H.R. acknowledges support from the West Virginia Higher Education Policy Commission through the Research Challenge Grant Program (Award No. RCG 23-007, 2022). J.S. was supported by the European Research Council (ERC) under the European Union’s Horizon 2020 research and innovation program project HERO Grant Agreement No. 810451. Computational resources were also provided by ETH Zurich and by the Swiss National Supercomputing Center (CSCS) under project ids s1128 and s1273.

\bibliography{bibliography}
\end{document}